\documentclass[10pt, conference, letterpaper]{IEEEtran}
\IEEEoverridecommandlockouts
\usepackage{cite}
\usepackage{amsmath,amssymb,amsfonts}
\usepackage{graphicx}
\usepackage{subfigure} 
\usepackage{textcomp}
\usepackage{color,xcolor}
\usepackage[caption=false]{subfig}
\usepackage{algorithm}
\usepackage{algpseudocode}
\usepackage{mathrsfs}
\usepackage{bm}
\usepackage[ruled,algo2e,linesnumbered]{algorithm2e}
\usepackage{multirow}
\usepackage{booktabs}
\usepackage{array}

\def\BibTeX{{\rm B\kern-.05em{\sc i\kern-.025em b}\kern-.08em
    T\kern-.1667em\lower.7ex\hbox{E}\kern-.125emX}}

\begin{document}

\title{Detecting Adversarial Spectrum Attacks via Distance to Decision Boundary Statistics}
\author{
\IEEEauthorblockN{Wenwei Zhao\IEEEauthorrefmark{1}, Xiaowen Li\IEEEauthorrefmark{2}, Shangqing Zhao\IEEEauthorrefmark{3}, Jie Xu\IEEEauthorrefmark{4}, Yao Liu\IEEEauthorrefmark{2}, Zhuo Lu\IEEEauthorrefmark{1}\\}
\IEEEauthorblockA{\IEEEauthorrefmark{1}Department of Electrical Engineering, University of South Florida}
\IEEEauthorblockA{\IEEEauthorrefmark{2}Department of Computer Science and Engineering, University of South Florida}
\IEEEauthorblockA{\IEEEauthorrefmark{3}School of Computer Science, University of Oklahoma}
\IEEEauthorblockA{\IEEEauthorrefmark{4}Department of Electrical and Computer Engineering, University of Miami}
}

\maketitle

\begin{abstract}
Machine learning has been adopted for efficient cooperative spectrum sensing. However, it incurs an additional security risk due to attacks leveraging adversarial machine learning to create malicious spectrum sensing values to deceive the fusion center, called adversarial spectrum attacks. In this paper, we propose an efficient framework for detecting adversarial spectrum attacks. Our design leverages the concept of the distance to the decision boundary (DDB) observed at the fusion center and compares the training and testing DDB distributions to identify adversarial spectrum attacks. We create a computationally efficient way to compute the DDB for machine learning based spectrum sensing systems. Experimental results based on realistic spectrum data show that our method, under typical settings, achieves a high detection rate of up to 99\% and maintains a low false alarm rate of less than 1\%. In addition, our method to compute the DDB based on spectrum data achieves 54\%--64\% improvements in computational efficiency over existing distance calculation methods. The proposed DDB-based detection framework offers a practical and efficient solution for identifying malicious sensing values created by adversarial spectrum attacks.

\end{abstract}
\begin{IEEEkeywords}
Spectrum sensing, adversarial machine learning, attack detection
\end{IEEEkeywords}

\section{Introduction}
With the ra pid development of wireless communication technology, spectrum resources are strained \cite{kolodzy2004dynamic,chiang2007quantitative}. Cooperative spectrum sensing can help improve the utilization of these resources by allowing multiple nodes to collect sensing data and report it to the fusion center \cite{peh2010cooperative,atapattu2011energy}. The fusion center then uses the sensing data to determine whether a primary channel is occupied or not such that unoccupied channels may be utilized while minimizing interference to primary users on occupied channels. However, it is possible for attackers to manipulate the sensing data in an attempt to deceive the fusion center into making incorrect decisions about channel availability. Attacks against cooperative spectrum sensing can take various forms, depending on the specific design of the system and the goals of the attacker. For example, an attacker could try to manipulate the sensing process by altering the signals received by the nodes \cite{luo2020attackers}, or by introducing false signals into the network~\cite{li2019speed}. An attacker could also try to interfere with the communication between nodes, in an attempt to disrupt the cooperative sensing process \cite{yimobicom}. These attacks can have serious consequences, including disruption to primary user communication and low spectrum utilization.

The proliferation of machine learning techniques has facilitated some recent works\cite{liu2019deep, xie2019activity} to utilize machine learning for automatically learning features from data to effectively accomplish spectrum sensing tasks. Machine learning-based defenses \cite{wang2018primary, rajasegarar2015pattern} have been proposed to counter different spectrum sensing attacks. However, this technological advancement has also presented new challenges as attackers have devised ways to exploit vulnerabilities in these systems. Recent studies \cite{luo2020attackers, liu2022attacking} have proposed to launch adversarial attacks against fusion center models from the perspective of machine learning. An adversarial attack is a type of attack on a machine learning model in which the attacker intentionally provides input to the model that is designed to cause the model to make a mistake \cite{vorobeychik2018adversarial} with minimum data change. It has been shown \cite{luo2020attackers} that such attacks can effectively beat traditional methods that defend against spectrum data falsification. We call such attacks {\it adversarial spectrum attacks}. 

To protect cooperative spectrum sensing against adversarial spectrum attacks, several methods have been proposed recently in the literature. The work in \cite{zheng2021primary} proposes an effective defense based on autoencoder against adversarial attacks without affecting the performance of the model, but it runs on static datasets and does not account for any data variations or dynamics commonly existing in wireless environments. The work in \cite{luo2020attackers} provides a method to mitigate harmful samples based on limiting the influence of each individual node on the fusion center's decision, but incurs a high computational cost and degrades the underlying sensing performance with the presence of an attack. 


Due to the limitations of current approaches in spectrum sensing scenarios, our focus in this paper is on creating an effective defense to detect the presence of adversarial machine learning against cooperative spectrum sensing. The essential idea of our defense is to leverage the concept of the distance to the decision boundary (DDB) \cite{mickisch2020understanding}. Decision boundaries are the boundaries that separate the different classes of data. The decision boundary in machine learning is determined during the training phase when the model learns to distinguish between the different classes based on the input features and their corresponding labels. Generally, the attackers aim to create subtle perturbations to deceive the classifier, which makes the data closer to a decision boundary. Based on this observation, it is possible to build a defense based on the DDB statistics of sensing data over time. Even though the adversarial spectrum attacker can minimize the sensing data change to flip the fusion center's decision, the resulting data will incur a different DDB compared with the original data. Thus, we can use a distance statistic metric to measure the similarity between the original training data and the testing data to indicate the presence of an adversarial spectrum attack at the fusion center. There are two major components in our proposed detection mechanism: i) computing the DDB for an input of spectrum sensing data and ii) forming a DDB statistic over time to measure the similarity. 

Finding and computing the DDB is essential in our approach. Typically, we need to determine the exact location of the decision boundary. However, in deep neural networks, locating the decision boundary can be challenging \cite{karimi2019characterizing, he2018decision}. The DeepFool method \cite{moosavi2016deepfool} proposes a way to compute the DDB based on a given model by iteratively perturbing input samples and checking the classifier's output until samples are misclassified. Other methods like Broyden-Fletcher-Goldfarb-Shanno (LBFGS) \cite{szegedy2013intriguing}, the Carlini and Wagner method (C\&W) \cite{carlini2017towards}, and the Decoupled Direction and Norm (DDN)\cite{rony2018decoupling} all aim to generate adversarial examples using different approximations to find the shortest distance. They usually require a substantial number of iterations and gradient calculations, therefore incurring a high computational cost for the fusion center to make a timely decision toward efficiently using spectrum resources. We propose to find the shortest distance by doing a binary search for data points along the direction perpendicular to a linear decision surface approximation for spectrum sensing data until we find the data point lies on the decision boundary. This not only mitigates the computational complexity of the method but also makes the finding of the DDB more straightforward for spectrum data classification. 

After we get the DDBs in a sequence of spectrum sensing data inputs, we adopt the Kolmogorov-Smirnov test \cite{massey1951kolmogorov} to compare the DDB distribution of the data inputs with that of the training data. We collect realistic spectrum data and use experiments to show that the proposed detection is able to effectively detect the presence of adversarial spectrum attacks with typically 97\% -- 99\% detection rates while maintaining low false alarms no more than 1\% under the cooperative spectrum sensing scenarios with various settings. Our main contributions to this paper are as follows. 


 
In summary, our major contributions are as follows. 1) We propose an attack detection framework that leverages the metric of DDB to detect the presence of adversarial spectrum attacks. 2) We create a new method to locate the decision boundary and compute the DDB in spectrum sensing applications. The new method is more computationally efficient than existing methods adopted in the machine learning community. 3) The experimental results based on realistic spectrum data illustrate the efficiency and effectiveness of our proposed DDB-based attack detection compared to existing approaches.

The rest of the paper is organized as follows. In Section~\ref{Sec:Model}, we present the system model and the threat model in the paper. In Section~\ref{Sec:Detection}, we describe in detail the calculation of DDB and how we leverage the DDB to detect the presence of adversarial spectrum attacks. Next, we conduct experimental evaluations and discuss the results in Section~\ref{Sec:Evaluations}. Finally, we review the related work in Section~\ref{Sec:RW} and conclude this paper in Section~\ref{Sec:Con}.

\section{System and Threat Models}\label{Sec:Model}
In this section, we describe the system models, threat models and state our research problem. 

\subsection{System Model}
We consider a cooperative spectrum sensing wireless network, in which there are $n$ sensing nodes and one fusion center. At the $i$-timeslot, each sensing node first senses the energy level of the wireless channel. We denote all sensed values by a column vector $\mathbf x_i = [x_{i,1},x_{i,2},...,x_{i,n}]^\top \in \mathbb{R}^{n\times1}$, where $x_{i,j}$ ($j\in[1,n]$) is the $j$-th node's sensed energy value; and $\mathbb{R}^{n\times1}$ denotes the $n$-dimensional real space. Then, all sensing nodes report $\mathbf x_i$ to a sensing data classifier $f$ at the fusion center. The classifier $f$ first uses two prediction functions $f_1:\mathbb{R}^{n\times1}\rightarrow\mathbb{R}$ and $f_0:\mathbb{R}^{n\times1}\rightarrow\mathbb{R}$ to compute the prediction scores for labels 1 and 0, denoting the channel available and unavailable, respectively. Then, $f$ outputs the label with the higher prediction score as the sensing decision $y_i$ for the $i$-th timeslot. 

\subsection{Threat Model}
Due to the distributed nature of cooperative spectrum sensing, the fusion center's decision relies on the individual node's report. However, it is not always guaranteed \cite{atapattu2011energy} that each node in the network will not be compromised or behave in a selfish or malicious way. There are existing efforts in the literature on studying various attack and defense strategies in cooperative spectrum sensing. Such attacks are usually called Spectrum Sensing Data Falsification (SSDF) attacks~\cite{fragkiadakis2012survey}. Traditionally, to detect the presence of SSDF, there are several major design categories: (i) reputation based detection~\cite{chen2008robust, nguyen2009robust}, (ii) machine learning based detection \cite{zhang2022speckriging, li2019scaling}, (iii) cross-correlation based detection~\cite{xu2009double}.

\begin{figure}[!htbp]
  \centering
  \includegraphics[width=0.9\linewidth]{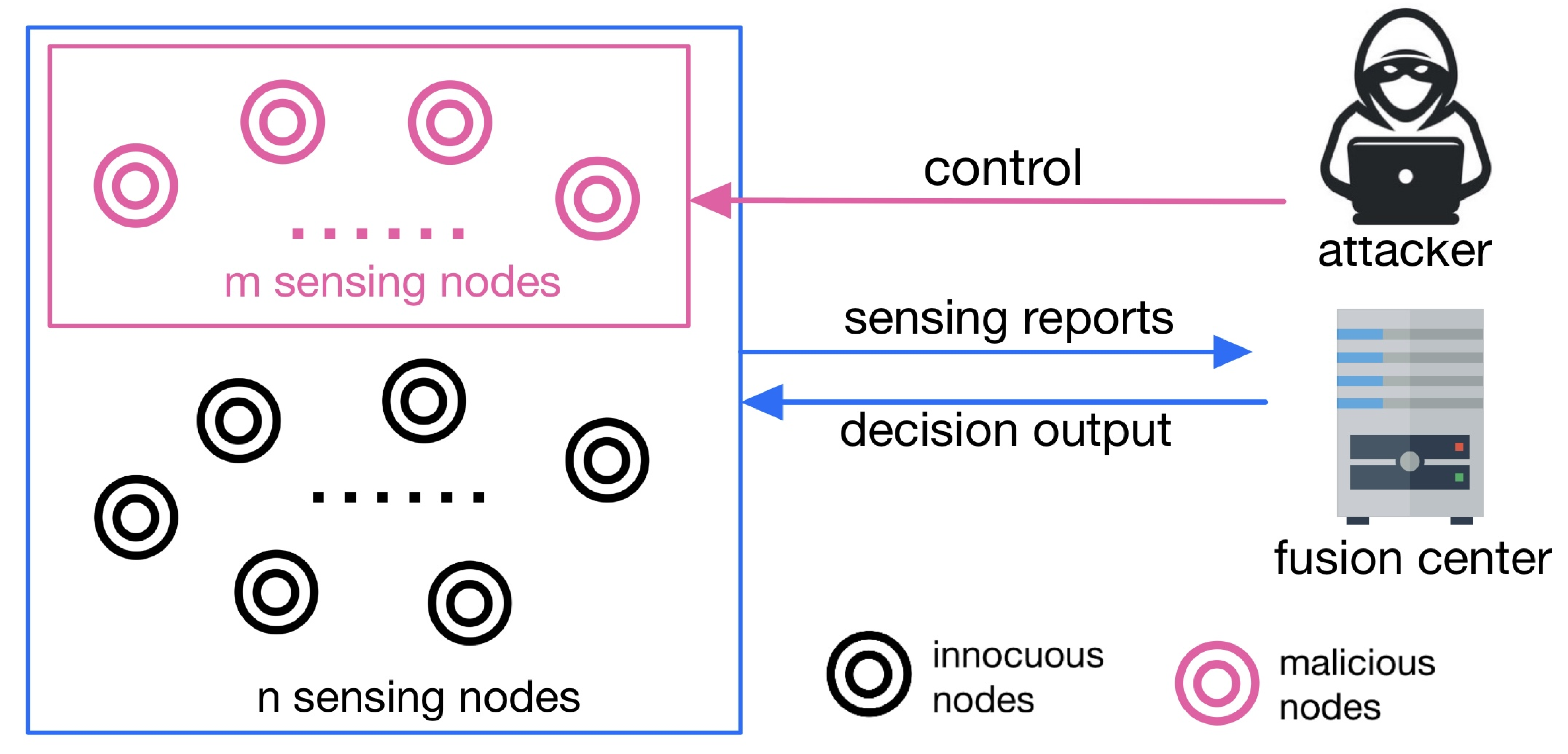}
  \caption{Cooperative sensing scenario with malicious nodes controlled by an attacker.}
  \label{fig: system}
\end{figure}

Recently, due to the advancement of adversarial machine learning, a new type of attack \cite{luo2020attackers} has been proposed against cooperative spectrum sensing. For example, when an attacker can control $m<n$ nodes in the network \cite{luo2020attackers} (without loss of generality, assuming nodes 1-$m$ are compromised), as shown in Fig.~\ref{fig: system}, the attacker can consider the decision function $f$ of the fusion center as a black box model with sensing data $\mathbf x_i$ as the input and decision $y_i$ as the output. In this way, the attacker can observe a series of inputs and outputs in the network and build its own machine learning model to predict the decision of the fusion center. Then, the attacker can create adversarial examples at the $m$ nodes that it controls \cite{chen2017cooperative} as falsified sensing reports to fool the fusion center to make an incorrect decision.

Mathematically, the attacker can create a new falsified data vector $\mathbf x'_i = \mathbf x_i+\mathbf{\delta}_i$ by building the surrogate model $S_i$ at timeslot~$i$ and find the perturbation vector $\mathbf{\delta}_i$ that satisfies:
\begin{eqnarray}
  \text{Objective:} &  \underset{\mathbf\delta_i \in \mathbb{R}^n}{\arg\min} {\Vert \mathbf\delta_i \Vert}_2 \\
  \text{s.t.} & S_i(\mathbf x'_i) \neq S_i(\mathbf x_i)
\end{eqnarray}

It has been shown in \cite{luo2020attackers, luo2021low} that such attacks can greatly degrade the performance of cooperative spectrum sensing even under traditional defense methods. We call such attacks leveraging adversarial machine learning to create falsified sensing data as {\it adversarial spectrum attacks}.

\subsection{Problem Statement}
The adversarial spectrum attack is an emerging security threat against cooperative spectrum sensing. Yet, there is no systematic study to detect such an attack in a cooperative sensing network. Some existing methods are limited in certain application scenarios and cannot be directly applied. For example, the autoencoder-based defense method \cite{zheng2021primary} does not take into account any data variations or dynamics in wireless environments, and the randomized smoothing \cite{kim2021channel} may decrease a model's prediction accuracy. The efforts have also been made in \cite{luo2020attackers} to mitigate the impact of adversarial spectrum attacks by limiting the influence of individual nodes on the fusion center's global decision. However, this leads to a non-negligible penalty on the sensing performance. 

As a result, we consider the problem of adversarial spectrum attacks in this paper. We aim to propose a new detection method to accurately detect the presence of adversarial spectrum attacks in a cooperative spectrum sensing network. 

\section{Detection Leveraging DDB}~\label{Sec:Detection}
In this section, we first define the decision boundary and the DDB and briefly introduce existing methods to find the DDB. Then, we propose our method to find the DDB for cooperative spectrum sensing applications. 

\subsection{Decision Boundary and DDB}

\subsubsection{Definitions}
Some current studies \cite{mickisch2020understanding, madry2017towards} have pointed out that subtle changes in the data will lead to significant changes in the distance between data and the decision boundary of a decision model. 



The decision boundary and the DDB are defined based on the prediction functions $\{f_y\}_{y\in\{0, 1\}}$ in the classifier at the fusion center. The classification decision is given by 
\begin{equation}
\widehat{y}(\mathbf{x}) = \arg\max\limits_{y} f_y(\mathbf{x}).
\end{equation}

The decision boundary $\mathcal{D} \subset\mathbb{R}^{n \times 1}$ for $y\in \{0, 1\}$ is defined as 
\begin{equation}
    \mathcal{D}:=\{\mathbf{x} | f_0(\mathbf{x} ) = f_1(\mathbf{x})\} \label{eq:boundary}.
 \end{equation}

Then, for a given data vector $\mathbf x$, we use the $\ell_2$-norm to define its distance to the decision boundary 
\begin{eqnarray}
d(\mathbf{x}) = \min\limits_{\mathbf{\delta} \in \mathbb{R}^n} {\Vert \mathbf{\delta} \Vert}_2 \quad \mathrm{ s.t. } \quad f_0(\mathbf{x}+\mathbf{\delta}) = f_1(\mathbf{x}+\mathbf{\delta})\label{eq:mini_distance}.
 \end{eqnarray}


\subsubsection{Finding Decision Boundary and DDB}
Recently, there have been increasing efforts in the machine learning community to propose methods to investigate the decision boundary of a neural network model \cite{mickisch2020understanding, nar2019cross, he2018decision} generate particularly small adversarial perturbations for the data that keep it at the decision boundary approximation and considered adversarial examples as a type of DDB. Commonly used methods to find the DDB include DeepFool~\cite{moosavi2016deepfool}, Limited memory Broyden-Fletcher-Goldfarb-Shanno (LBFGS)~\cite{szegedy2013intriguing}, Carlini and Wagner (C\&W)~\cite{carlini2017towards} methods. 






\noindent{\bf DeepFool:} The work of \cite{mickisch2020understanding} takes the approximation in \cite{elsayed2018large} based on DeepFool to find the distance of a point to the decision boundary. To compute the distance $d(\mathbf{x})$ in \eqref{eq:boundary}, the closed-form of perturbation $\delta_{\{k\}}$ at the $k^{th}$ iteration can be written as
\begin{equation}
  \mathbf \delta_{\{k\}} (\mathbf{x}) = \frac{|f_0(\mathbf{x}_{\{k\}})-f_1(\mathbf{x}_{\{k\}})|}{{\Vert \nabla f_0(\mathbf{x}_{\{k\}})-\nabla f_1(\mathbf{x}_{\{k\}}) \Vert}_2^2} \, D_{\nabla}(\mathbf{x}_{\{k\}})
\end{equation}
where $D_{\nabla}(\mathbf{x}_{\{k\}}) = |\nabla f_0(\mathbf{x}_{\{k\}})-\nabla f_1(\mathbf{x}_{\{k\}})|$ and $\mathbf{x}_{\{k+1\}}=\mathbf{x}_{\{k\}}+ \mathbf \delta_{\{k\}} (\mathbf{x})$. When the iteration stops at the decision boundary after $K$ iterations, and the perturbation is given by 
$
\delta(\mathbf{x}) = \sum\limits_{k=0}^{K-1} \delta_{\{k\}} (\mathbf{x})
$, and we can get the DDB $d(\mathbf x) = {\Vert \mathbf \delta(\mathbf x) \Vert}_2$.

\noindent{\bf LBFGS:} A box-constrained optimizer is used in \cite{szegedy2013intriguing} to minimize the perturbation and approximate \eqref{eq:mini_distance} as 
\begin{eqnarray}
\min\limits_{\mathbf{\delta} \in \mathbb{R}^n} C{\Vert \mathbf{\delta} \Vert}_2 + {\rm loss}_{f} (\mathbf{x} + \mathbf{\delta}, y_{\rm target})\\
\mathrm{ s.t. }\mathbf{-M} 	\leq \mathbf{x}+\mathbf{\delta} \leq \mathbf{M},
\end{eqnarray}
where ${\rm loss}_{f} (\mathbf{x} + \mathbf{\delta}, y_{\rm target})$ is the loss function, $y_{\rm target}$ is the targeted label of the adversarial example, $\mathbf{-M}$ and $\mathbf{M}$ constrain the range of the adversarial example. A line search is applied to update the constant $C$ to optimize the function.

\noindent{\bf C\&W:} \cite{carlini2017towards} proposed to change variables by using the tanh function instead of using a box-constrained optimizer via alternating \eqref{eq:mini_distance} with the following equation:
\begin{eqnarray}
\label{eq:cw_distance}
\min\limits_{\mathbf{\delta} \in \mathbb{R}^n} {\Vert \mathbf{\delta} \Vert}_2 + CF(\mathbf{x}+\delta)	\\ 
\mathrm{ s.t. }\mathbf{-M} \leq \mathbf{x}+\mathbf{\delta} \leq \mathbf{M},
\label{eq:cw_constraints}
\end{eqnarray}
where function $F(\mathbf{x}+\mathbf{\delta})$ is based on the best objective function and is defined as 
\begin{eqnarray}
    F(\mathbf{x}+\mathbf{\delta})= \max(\max\{f(\mathbf{x}+\mathbf{\delta})_{y_1}\} - f(\mathbf{x}+\mathbf{\delta})_{y_0}, \mathbf{\kappa})\\
    \mathrm{ s.t.}~y_0, y_1 \in \{0,1\}, ~y_0 \neq y_1
\end{eqnarray}
with $\mathbf{\kappa}$ used to control the confidence in the occurrence of misclassification. Because of the box constraints in \eqref{eq:cw_constraints}, C\&W introduced a new variable $\mathbf{\rho}$ and used tanh function \cite{mishkin2015all} to rewrite the perturbation $\mathbf{\delta}$ as 
\begin{equation}
    \mathbf{\delta} = \frac{1}{2}({\rm tanh}(\mathbf{\rho})+1) - \mathbf{x}
\end{equation}
It then applied the change-of-variables and optimize over $\mathbf{\rho}$ for \eqref{eq:cw_distance}.

\subsection{Proposed Method to Compute DDB for Spectrum Sensing}
\subsubsection{Motivation}
\begin{figure}[!htbp]
  \centering
  \includegraphics[width=0.8\linewidth]{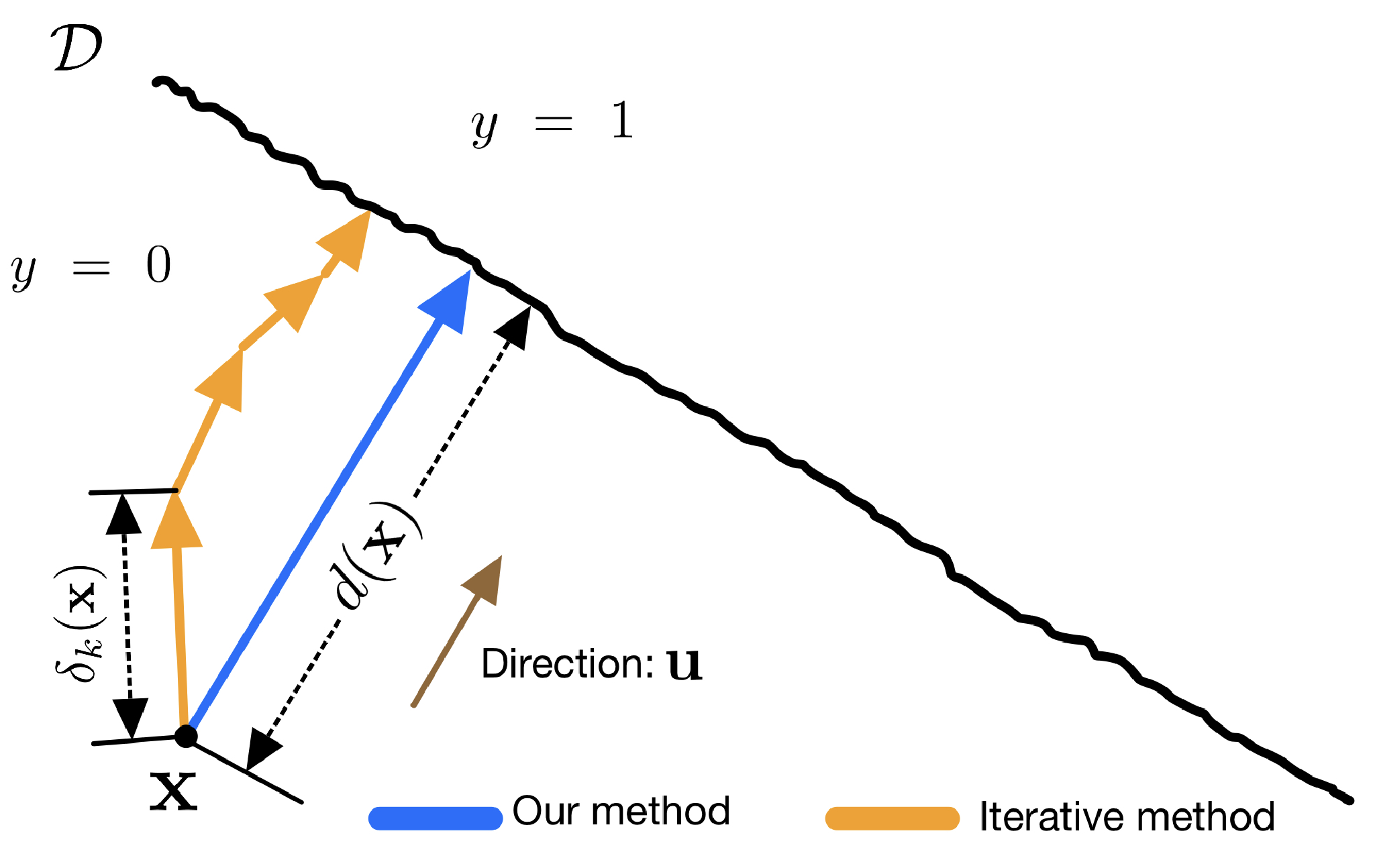}
  \caption{Our idea to find the direction vs an iterative method to find the DDB. }
  \label{fig: distance}
\end{figure}

DeepFool, LBFGS, and C\&W all adopt an iterative procedure to estimate the distance between a given data point and the decision boundary but have high computational complexity. In order to make DDB-based attack detection more effective, we propose a simpler and faster method to calculate the DDB. Our observation is that in spectrum sensing, when the signal energy level increases, the wireless channel is more likely considered occupied; similarly, when the energy level decreases, the channel is more likely to be available. Thus, adding or deducting the value of the energy level should always flip the decision of the fusion center. By increasing or decreasing the values in $\mathbf x_i$ at the $i$-th timeslot along a certain direction $\mathbf{u}$, we can find the shortest distance from the data point to the decision boundary of whether the channel is available or not. Fig.~\ref{fig: distance} gives illustrative examples to compare an iterative method with our idea of finding the DDB. 

Then, our goal becomes designing a computationally efficient method to find the direction along which we can estimate the DDB for spectrum sensing classification. To this end, we first notice that using machine learning for cooperative spectrum sensing is commonly motivated by the fact that sensing data can be unreliably collected and vary over time. Thus, machine learning becomes an efficient way to characterize such data without explicitly assuming particular sensing data models \cite{luo2020attackers}. However, the advantage of assuming sensing data distributions is that an optimal solution in terms of maximizing a performance metric can be developed. When a machine learning classifier is sufficiently trained by the sensing data, it should achieve a performance close to the optimal solution. Hence, our idea is to use the theoretically optimal solution as a guide for estimating the direction that we should use to compute the DDB for a machine learning classifier. 

\subsubsection{Finding the Direction to Compute DDB}
We adopt the theoretical framework of the likelihood ratio test (LRT) based on the Neyman-Pearson lemma \cite{liu2019deep} to establish the decision boundary between sensing output labels 1 (channel available) and 0 (not available). Once the decision boundary is established, we use the direction from $\mathbf x_i$ perpendicular to the theoretical boundary to estimate the shortest distance to the actual boundary in the classifier. 

We assume that the fusion center has a full view of all signals from sensing nodes. Denote by $\mathbf{v}_i(t) \in \mathbb C^{n \times 1}$ the vector of the $t$-th received signal samples ($t\in\{1,, 2, \cdots, T\}$) at all nodes at the $i$-th timeslot, where $\mathbb C^{n \times 1}$ denotes the $n$-dimentional complex space. We can write $\mathbf{v}_i(t) = \mathbf{s}_i(t) + \mathbf{\eta}_i(t)$, where $\mathbf{s}_i(t)$ is the deterministic complex signal sent by a primary user and $\mathbf{\eta}_i(t)$ is the Gaussian noise; and 
\begin{eqnarray}
  \mathbf{v}_i(t)=\left\{
    \begin{array}{ll}
      \mathbf{s}_i(t) + \mathbf{\eta}_i(t), & \text{channel unavailable}\\
      \mathbf{\eta}_i(t), & \text{channel available}.
    \end{array} \right. \label{eq:singal_values}
\end{eqnarray}

\subsubsection{Analysis and Design}
The signal power sensed by node~$j$ at timeslot~$i$ can be denoted as $x_{ij} = \frac{1}{T} \sum_{t=1}^{T}{|v_{ij}(t)|}^2$, where $v_{ij}(t)$ is the $j$-th element in $\mathbf v_{i}(t)$. In cooperative spectrum sensing, the fusion center collects the signal power value instead of the signal samples from each node and makes a decision based on the fusion rule. Based on \eqref{eq:singal_values}, the distributions of the signal power when the channel is available (hypothesis $\mathscr{H}_1$) and unavailable (hypothesis $\mathscr{H}_0$) are both Gamma distributions with same shape parameter $T$, but different scale parameters $\beta_{ij}^0$ and $\beta_{ij}^1$. 
\begin{eqnarray}
  \label{eq:H0}
  \mathscr{H}_0 &:& x_j \sim \Gamma(T, \beta_{ij}^0),\\
  \mathscr{H}_1 &:& x_j \sim \Gamma(T, \beta_{ij}^1).
  \label{eq:H1}
\end{eqnarray}
After mathematical manipulations, the probability density functions of \eqref{eq:H0} and \eqref{eq:H1} can be expressed as 
\begin{eqnarray}
  p(x_j | \mathscr{H}_0) = \frac{x_j^{T-1}e^{-\frac{x_j}{\beta_{ij}^0}}}{{(\beta_{ij}^0)}^T \Gamma(T)},\\
  p(x_j | \mathscr{H}_1) = \frac{x_j^{T-1}e^{-\frac{x_j}{\beta_{ij}^1}}}{{(\beta_{ij}^1)}^T \Gamma(T)}.
\end{eqnarray}

Then, according to LRT, we can write 
\begin{eqnarray}
   \Lambda = \frac{p(\mathbf{x}_i | \mathscr{H}_1)}{p(\mathbf{x}_i | \mathscr{H}_0)} =  \frac{\prod \limits_{j=1}^n p(x_{ij} | \mathscr{H}_1)}{\prod\limits_{j=1}^n p(x_{ij} | \mathscr{H}_0)}\label{eq:LRT} \mathop{\gtrless}\limits_{\mathscr H_{0}}^{\mathscr H_{1}}\gamma,
\end{eqnarray}
where $\gamma$ is a given threshold for detection. After mathematical manipulations, \eqref{eq:LRT} can be expressed as 
\begin{eqnarray}
    \sum_{j = 1}^{n}\left(\frac{1}{\beta_{ij}^0} - \frac{1}{\beta_{ij}^1}\right)x_{ij} \mathop{\gtrless}\limits_{\mathscr H_{0}}^{\mathscr H_{1}} \ln(\gamma) - T \sum_{j=1}^n \ln\left(\frac{\beta_{ij}^0}{\beta_{ij}^1}\right) 
    \label{eq:DB}
\end{eqnarray}

It can be seen that the decision boundary of the hypotheses $\mathscr{H}_0$ and $\mathscr{H}_1$ is linear, and can be expressed as $\mathcal{D}= \{\mathbf{x}: \mathbf{w}^\top\mathbf{x} +b = 0\}$, where $\mathbf{w} = [(\frac{1}{\beta_{1}^0} - \frac{1}{\beta_{1}^1}), (\frac{1}{\beta_{2}^0} - \frac{1}{\beta_{2}^1}),..., (\frac{1}{\beta_{n}^0} - \frac{1}{\beta_{n}^1})]$, where $\beta_j^0 = \beta_{ij}^0$ and $\beta_j^1 = \beta_{ij}^1$ for all timeslots. Thus, for a given sensing vector $\mathbf{x_i}$, the vector $\mathbf{w}$ denotes the direction of its shortest path to the decision boundary $\mathcal D$, and its DDB can be given by
\begin{eqnarray}
d(\mathbf{x}_i) = \min\limits_{\delta \in \mathbb{R}^{n \times 1}} {\Vert \mathbf{\delta} \Vert}_2 = \frac{|f(\mathbf{x}_i)|}{{\Vert \mathbf{w} \Vert}_2} \label{eq:distance} \\
    \mathrm{ s.t. }~ f_0(\mathbf{x}_i+\mathbf{\delta}) = f_1(\mathbf{x}_i+\mathbf{\delta})
\end{eqnarray}
where $f$ is the prediction function of the fusion center, $f_0$ and $f_1$ are the prediction scores for classes 0 and 1 of classifier $f$, respectively. 

A binary search strategy can be used to efficiently solve \eqref{eq:distance}. More specifically, to find the distance from the sensing vector $\mathbf x_i$ to the decision boundary $\mathcal D$, we create a new vector along the search direction $\mathbf{u} = -\frac{\mathbf{w}}{{\Vert \mathbf{w} \Vert}_2}$ as $\mathbf x_i + \epsilon \mathbf{u}$, where $\epsilon$ is large enough such that the new vector flips the fusion center's decision. Then, the binary search is used to find the boundary point from $\mathbf x_i$ to $\mathbf x_i + \epsilon \mathbf{u}$ along the direction $\mathbf{u}$. The distance is then calculated as the Euclidean distance between $\mathbf x_i$ and the boundary point. The pseudocode of this algorithm is shown in Algorithm~\ref{alg:1}. 

\begin{algorithm2e}[t]
  \SetAlgoLined
  \KwData{Sensing data $\mathbf{x_i}$, initial step length $\epsilon$, direction $\mathbf{u}$, stop threshold $\xi$;}
  \KwResult{DDB $\mathbf{\delta_i}$\;}
  \uIf {$y(\mathbf{x_i}) = 0$} {
    \While {$y(\mathbf{x_i} + \epsilon\mathbf{u}) = 0$}{
      $\epsilon = 2\epsilon$;
    }
    $\mathbf{x_l}\leftarrow \mathbf{x_i}$, $\mathbf{x_r}\leftarrow \mathbf{x_i}+\epsilon\mathbf{u}$\;
    }\uElseIf {$y(\mathbf{x_i}) = 1$} {
      \While {$y(\mathbf{x_i} - \epsilon\mathbf{u}) = 1$}{
        $\epsilon = 2\epsilon$;
      }
    $\mathbf{x_l}\leftarrow \mathbf{x_i}-\epsilon\mathbf{u}$, $\mathbf{x_r}\leftarrow \mathbf{x_i}$\;
    }\Else{
    \KwRet{ ``False"\;}
    }
  \Repeat{$|\mathbf{x_l}-\mathbf{x_r}| \leq \xi$}{
    $\mathbf{x}^{\prime}\leftarrow \frac{\mathbf{x_l}+\mathbf{x_r}}{2}$\;
    
    \uIf {$y(\mathbf{x_i}) = 0$} {
    $\mathbf{x_l} \leftarrow \mathbf{x}^{\prime}$\;
    }\uElseIf {$y(\mathbf{x_i}) = 1$} {
    $\mathbf{x_r}\leftarrow \mathbf{x}^{\prime}$\;
    }\Else{
    \KwRet{ ``False"\;}
    }
    }    
  \KwRet{$\mathbf{\delta_i} = |\mathbf{x}^{\prime}-\mathbf{x}|$\;}
  \caption{Pseudocode of DDB Algorithm}\label{alg:1}
\end{algorithm2e}

\subsection{Detection Method}
After computing the DDB, we can collect the DDB for each training sample during the training process and obtain the distribution of the DDB from training. This distribution will serve as the ground truth DDB distribution. Our intuition for attack detection is that the attacks based on adversarial machine learning always aim to minimize the data perturbation to just go across the decision boundary of a classifier. Thus, if we measure the DDB distribution of testing data under the adversarial spectrum attacks, it should be quite different from the training distribution. This becomes the basis for attack detection by comparing the training and testing DDB distributions. 

As a result, we use the consistency of the distributions between the training and testing DDB sets to detect whether there exists an adversarial spectrum attack against the fusion center. We choose the Kolmogorov-Smirnov (K-S) test\cite{massey1951kolmogorov}, which is commonly used as a nonparametric test to analyze whether two sets of data are different especially when the sample sizes in the two sets are relatively small. 

Given a training DDB set with size $a_1$ and a testing DDB set with size $a_2$, we calculate their respectively Cumulative Distribution Functions (CDFs) $F_{\text{train},a_1}(\delta)$ and $F_{\text{test},a_2}(\delta)$. The maximum distance $d_{\text{KS}}$ of the K-S statistic is $d_{\text{KS}} = \mathop{\max}\limits_{\delta}|F_{\text{train},a_1}(\delta) - F_{\text{test},a_2}(\delta)|$. Consider the null and alternative hypotheses
\begin{eqnarray}
  \mathcal{H}_0 &:& \text{$x_\text{train}$ and $x_\text{test}$ are from same distribution}\\
  \mathcal{H}_1 &:& \text{$x_\text{train}$ and $x_\text{test}$ are from different distributions}
\end{eqnarray}
where $\mathcal{H}_0$ indicates that there is no attack and $\mathcal{H}_1$ means that an attack with manipulating spectrum data. According to the K-S test, we can write the decision rule as 
\begin{eqnarray}
  P_\text{value} \mathop{\lesseqgtr}\limits_{\mathcal{H}_{0}}^{\mathcal{H}_{1}}\alpha,
\end{eqnarray}
where 
\begin{equation}\label{eq:pvalue1}
  P_\text{value} = 2e^{-2d_{\text{KS}}^2 \frac{a_1 a_2}{a_1+ a_2}},
\end{equation}
and the decision threshold $\alpha$ is called the significance level. 

\subsection{Can Attack Bypass DDB-based Detection?}
Our attack detection is to first compute the DDBs of spectrum sensing data vectors, then compare the DDB distributions to indicate the presence of an attack. The motivation behind the detection is that adversarial sensing data vectors manipulated by an attacker should always have very small distances to the decision boundary, but benign sending data vectors do not. Is it possible for an attacker to create adversarial spectrum sensing data while maintaining the same DDB distribution to evade our DDB-based detection method? In the following, we show that the attacker cannot create such attacks because of a lack of information. 

In our threat model in Section \uppercase\expandafter{\romannumeral2}, the attacker can only access and control $m$ out of $n$ ($m < n$) nodes in the network. That means for sensing data vector $\mathbf{x_i} \in \mathbb{R}^{n\times1}$ at timeslot $i$, only $[x_{i,1},x_{i,2},...,x_{i,m}]^\top \in \mathbb{R}^{m\times1}$ can be changed by the attacker and $[x_{i,m+1},x_{i,m+2},...,{x_{i,m+(n-m-1)}},x_{i,n}]^\top \in \mathbb{R}^{n-m\times1}$ is always the same regardless attack or not. If the attacker wants to defeat the DDB-based defense, the attacker should generate malicious data that will be classified into a different label instead of the original label, and at the same time, the DDB of the malicious data should have the same distribution as the ground truth.

Suppose that the attacker generates an adversarial sensing data vector with a target DDB of $d_t$ from the vector to the decision boundary. According to \eqref{eq:distance}, the objective function can written as
\begin{eqnarray}
  d_t = \frac{|\mathbf{w}^\top\mathbf{x'_i} +b|}{{\Vert \mathbf{w} \Vert}_2}
  \label{eq:adv}
  \end{eqnarray}
where $\mathbf{x'_i}$ is the adversarial spectrum data sensing vector, and \eqref{eq:adv} can be further expressed as 
\begin{eqnarray}
  \sum_{j = 1}^{m}w_j x'_{ij} = \pm d_t {\Vert \mathbf{w} \Vert}_2 -\sum_{j = m+1}^{n}w_j x_{ij} - b
  \label{eq:adv2}
  \end{eqnarray}
To find such an adversarial vector $\mathbf{x'_i}$, the attacker must know all the parameters in \eqref{eq:adv2}, and get the values of $[x_{i,m+1},x_{i,m+2},...,x_{i,n}]$. However, without the information of remaining $n-m$ nodes, the attacker cannot get the $n-m$ values in vector $\mathbf{x_i}$ and $\mathbf{w}$. As a result, it is generally infeasible for the attacker to create an adversarial vector with a specifically targeted DDB.

\section{Experimental Evaluations}\label{Sec:Evaluations}
In this section, we present our experimental evaluations. We first introduce the experimental setup and spectrum data set collection, then discuss the evaluation results. 

\subsection{Evaluation Setups}
\subsubsection{Spectrum Sensing Scenario}
We set up a cooperative sensing network consisting of the data fusion center and $n=20$ sensing nodes, in which $m<10$ nodes are compromised and try to launch adversarial attacks against the fusion center. We use realistic spectrum sensing data for the evaluation. In particular, the data includes realistic white space TV signal strengths collected by reliable RTL-SDR TV dongles. The signal strengths on the TV channels were collected simultaneously at 20 different locations (interior and exterior of a building or different floors in the building) at a university campus. The sensing process at each location measured the average signal power over a 30-second time period as the sensing result for each time slot, as required by the Federal Communications Commission\cite{coase1959federal}. 

\subsubsection{Spectrum Data Fusion Center}
We implement a multi-layer perceptron (MLP) classifier at the fusion center to classify the spectrum sensing data into benign or malicious. We use 20,000 spectrum sensing data vectors as training data to train the classifier, which can achieve a high sensing accuracy of 99.94\% without attack. Then, in our experiment, we use additional 80,000 sensing data vectors to simulate the attacking scenario in which $m$ malicious nodes aim to manipulate their corresponding entries in the sensing data vector to fool the fusion center. We note that when there is no attack, the classifier at the fusion center can always correctly classify each of the 80,000 sensing data vectors. 

\subsubsection{Attack Detection and Evaluation Metrics} 
We deploy the DDB-based detector at the fusion center to detect the malicious behavior of the $m$ malicious nodes. We aim to evaluate how our proposed DDB calculation method compares with the DDB estimates by using DeepFool, LBFGS, and C\&W in the literature. Once we obtain the DDBs from different calculation methods, we always use the K-S test to compare the DDB distributions. We use the detection rate and false alarm rate as the evaluation metrics for the detection performance. The detection rate is the probability that the detector indicates an attack in the presence of an attack. The false alarm rate is the probability that the detector indicates an attack, but there is no actual attack in the system.



  

\subsubsection{Default Settings} During our evaluations, we adopt the following default settings and evaluate the scenario in the presence of the attack. The first step length of Binary Search is $\epsilon = 5$ and the threshold to stop searching is $\xi = 0.01$. The number of malicious nodes is $m=7$ and the attacker uses the Fast Gradient Sign Method (FGSM) \cite{goodfellow2014explaining} to generate adversarial spectrum data vectors. In the K-S test, the group size $a_2$ for attack detection is 25; and the threshold $\alpha$ of $P_\text{value}$ is set to 0.01, meaning that the distributions of ground truth data and test data are different at least with the $99\%$ confidence level in the detection. 

\subsection{Evaluation Results}


\subsubsection{Impact of Test Group Size on Attack Detection}
Generally, the more test data involved in the detection, the more accurate the attack detection should be. However, involving more data prolongs the attack detection time. Thus, we first evaluate the impact of the test group size $a_2$ on the attack detection performance. We choose 5, 10, 20, 25, 50, 80, 100, 200, and 400 as the group size to detect the presence of the adversarial spectrum attack. 

\begin{figure*}[htbp]
  \centering
	\begin{minipage}{0.31\linewidth}
		\centering
		\includegraphics[width=0.95\linewidth]{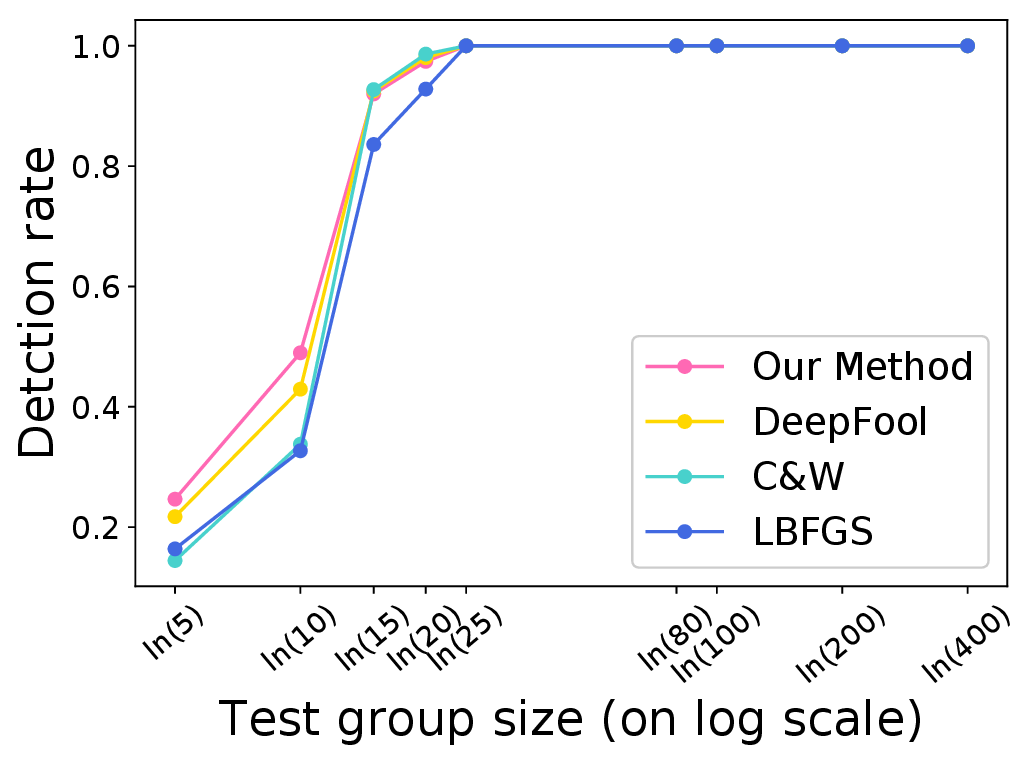}
    \vspace{-0.25cm}
    \caption{Detection rates under the different test group sizes.}
    \label{fig: e1_1}
	\end{minipage}
  \centering
	\begin{minipage}{0.31\linewidth}
		\centering
		\includegraphics[width=0.95\linewidth]{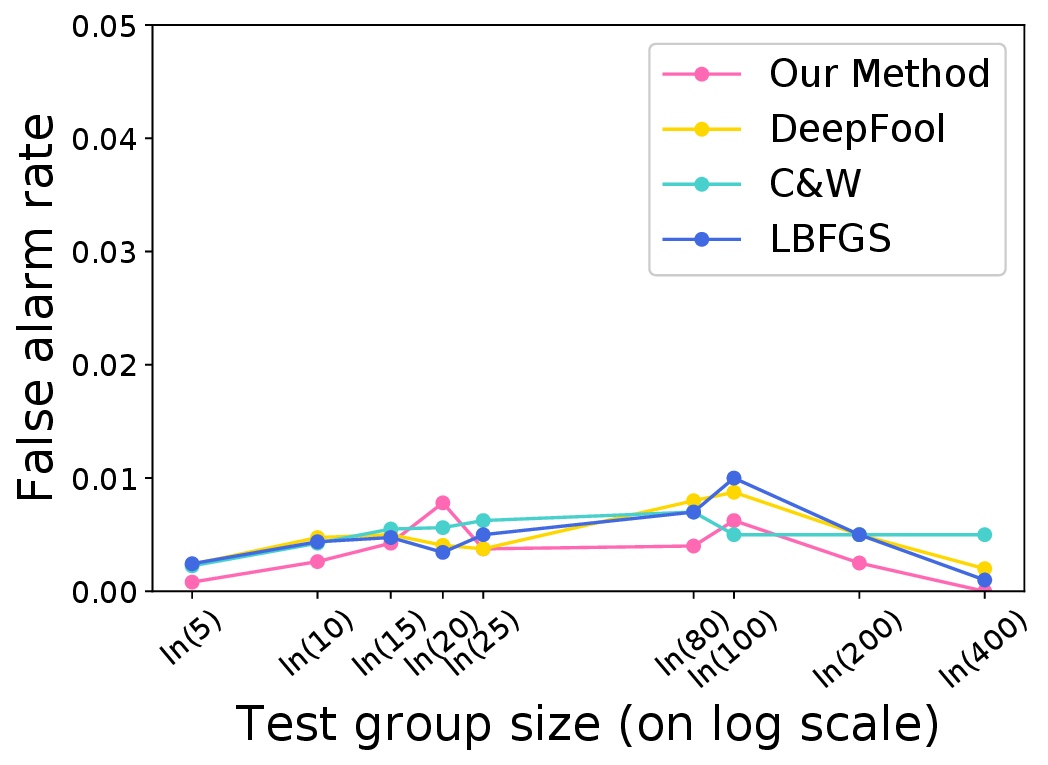}
    \vspace{-0.25cm}
		\caption{False alarms under the different sizes of test data.}
    \label{fig: e1_2}
	\end{minipage}
  \centering
  \begin{minipage}{0.31\linewidth}
    \centering
    \includegraphics[width=0.95\linewidth]{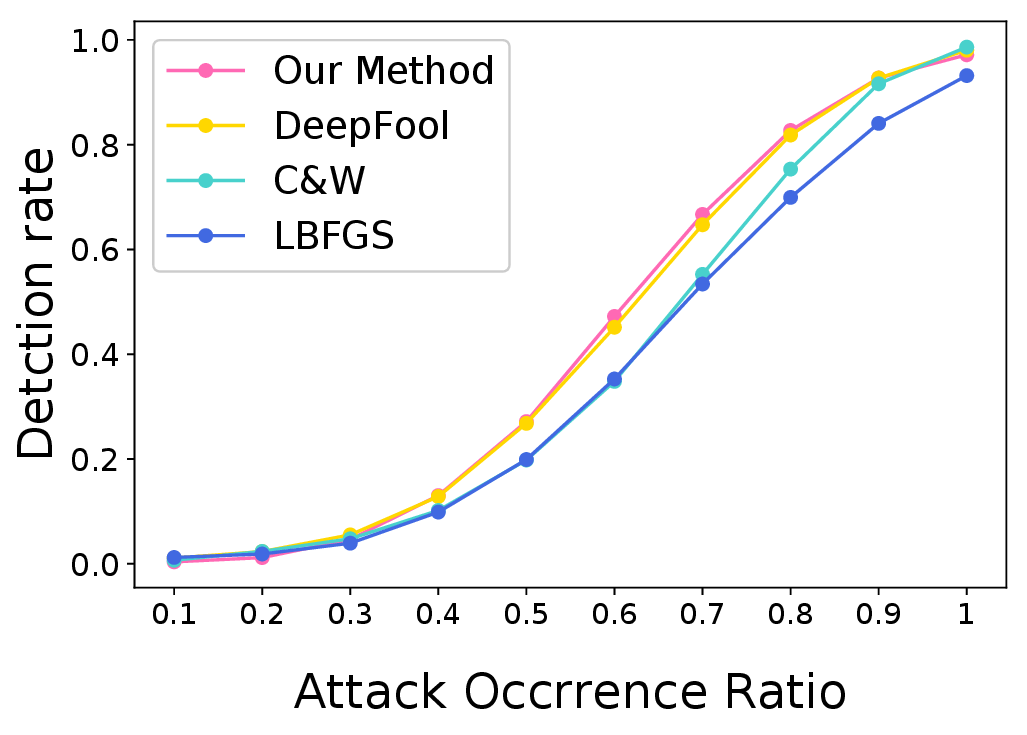}
    \vspace{-0.25cm}
    \caption{Detection rates under different attack occurrence ratios.}
    \label{fig: e7_3}
  \end{minipage}
\end{figure*}

Fig.~\ref{fig: e1_1} shows the detection rates of different methods as the group size increases from 5 to 400. First, we can see that improving the group size clearly increases the detection performance. Our proposed DDB method achieves approximately the same performance as existing methods including DeepFool, C\&W, and LBFGS. The attack detection rate approaches 100\% when the size of the group used for attack detection exceeds 20, indicating the relatively fast detection performance of 20 timeslot observation for each method. Fig.~\ref{fig: e1_2} shows the false alarm rates under the same conditions. We can see that all methods have low false alarm rates under 0.01. 

Overall, the results from Figs.~\ref{fig: e1_1} and~\ref{fig: e1_2} show that the DDB-based attack detection achieves good detection rates with low false alarm rates. In addition, our proposed DDB calculation method leads to the approximately same performance as DeepFool, C\&W, and LBFGS.

\begin{table*}[]
  \centering
  \caption{Detection rates under different attack occurrence ratios.}
  \renewcommand\arraystretch{0.5}
  \begin{tabular}{c|c|cccccccccc}
  \toprule
  \multirow{2}{*}{\textbf{Methodology}} & \multirow{2}{*}{\textbf{Group Size}} & \multicolumn{10}{c}{\textbf{Attack Occurrence Ratio}}                                                                                                                                                                                                                                                                                       \\ \cline{3-12} \rule{0pt}{8pt}
                                        &                                      & \multicolumn{1}{c|}{\textbf{0.1}} & \multicolumn{1}{c|}{\textbf{0.2}} & \multicolumn{1}{c|}{\textbf{0.3}} & \multicolumn{1}{c|}{\textbf{0.4}} & \multicolumn{1}{c|}{\textbf{0.5}} & \multicolumn{1}{c|}{\textbf{0.6}} & \multicolumn{1}{c|}{\textbf{0.7}} & \multicolumn{1}{c|}{\textbf{0.8}} & \multicolumn{1}{c|}{\textbf{0.9}} & \textbf{1.0} \\ \midrule
  \multirow{5}{*}{Our Method}           & 200                                  & \multicolumn{1}{c|}{0.0150}       & \multicolumn{1}{c|}{0.3800}       & \multicolumn{1}{c|}{0.9600}       & \multicolumn{1}{c|}{1.0000}       & \multicolumn{1}{c|}{1.0000}       & \multicolumn{1}{c|}{1.0000}       & \multicolumn{1}{c|}{1.0000}       & \multicolumn{1}{c|}{1.0000}       & \multicolumn{1}{c|}{1.0000}       & 1.0000       \\ \cline{2-12} \rule{0pt}{7pt}
                                        & 100                                  & \multicolumn{1}{c|}{0.0163}       & \multicolumn{1}{c|}{0.0950}       & \multicolumn{1}{c|}{0.4850}       & \multicolumn{1}{c|}{0.8763}       & \multicolumn{1}{c|}{0.9888}       & \multicolumn{1}{c|}{1.0000}       & \multicolumn{1}{c|}{1.0000}       & \multicolumn{1}{c|}{1.0000}       & \multicolumn{1}{c|}{1.0000}       & 1.0000       \\ \cline{2-12} \rule{0pt}{7pt}
                                        & 50                                   & \multicolumn{1}{c|}{0.0075}       & \multicolumn{1}{c|}{0.0244}       & \multicolumn{1}{c|}{0.1338}       & \multicolumn{1}{c|}{0.4094}       & \multicolumn{1}{c|}{0.7438}       & \multicolumn{1}{c|}{0.9056}       & \multicolumn{1}{c|}{0.9831}       & \multicolumn{1}{c|}{0.9981}       & \multicolumn{1}{c|}{1.0000}       & 1.0000       \\ \cline{2-12} \rule{0pt}{7pt}
                                        & 25                                   & \multicolumn{1}{c|}{0.0038}       & \multicolumn{1}{c|}{0.0119}       & \multicolumn{1}{c|}{0.0466}       & \multicolumn{1}{c|}{0.1303}       & \multicolumn{1}{c|}{0.2716}       & \multicolumn{1}{c|}{0.4722}       & \multicolumn{1}{c|}{0.6669}       & \multicolumn{1}{c|}{0.8272}       & \multicolumn{1}{c|}{0.9272}       & 0.9741       \\ \cline{2-12} \rule{0pt}{7pt}
                                        & 10                                   & \multicolumn{1}{c|}{0.0044}       & \multicolumn{1}{c|}{0.0076}       & \multicolumn{1}{c|}{0.0143}       & \multicolumn{1}{c|}{0.0334}       & \multicolumn{1}{c|}{0.0574}       & \multicolumn{1}{c|}{0.1051}       & \multicolumn{1}{c|}{0.1743}       & \multicolumn{1}{c|}{0.2626}       & \multicolumn{1}{c|}{0.3706}       & 0.4833       \\ \hline  \rule{0pt}{7pt}
  \multirow{5}{*}{DeepFool}             & 200                                  & \multicolumn{1}{c|}{0.0375}       & \multicolumn{1}{c|}{0.4050}       & \multicolumn{1}{c|}{0.9650}       & \multicolumn{1}{c|}{1.0000}       & \multicolumn{1}{c|}{1.0000}       & \multicolumn{1}{c|}{1.0000}       & \multicolumn{1}{c|}{1.0000}       & \multicolumn{1}{c|}{1.0000}       & \multicolumn{1}{c|}{1.0000}       & 1.0000       \\ \cline{2-12} \rule{0pt}{7pt}
                                        & 100                                  & \multicolumn{1}{c|}{0.0175}       & \multicolumn{1}{c|}{0.1063}       & \multicolumn{1}{c|}{0.5163}       & \multicolumn{1}{c|}{0.8988}       & \multicolumn{1}{c|}{0.9988}       & \multicolumn{1}{c|}{1.0000}       & \multicolumn{1}{c|}{1.0000}       & \multicolumn{1}{c|}{1.0000}       & \multicolumn{1}{c|}{1.0000}       & 1.0000       \\ \cline{2-12} \rule{0pt}{7pt}
                                        & 50                                   & \multicolumn{1}{c|}{0.0156}       & \multicolumn{1}{c|}{0.0406}       & \multicolumn{1}{c|}{0.1719}       & \multicolumn{1}{c|}{0.4231}       & \multicolumn{1}{c|}{0.7363}       & \multicolumn{1}{c|}{0.9206}       & \multicolumn{1}{c|}{0.9856}       & \multicolumn{1}{c|}{0.9981}       & \multicolumn{1}{c|}{1.0000}       & 1.0000       \\ \cline{2-12} \rule{0pt}{7pt}
                                        & 25                                   & \multicolumn{1}{c|}{0.0109}       & \multicolumn{1}{c|}{0.0234}       & \multicolumn{1}{c|}{0.0553}       & \multicolumn{1}{c|}{0.1284}       & \multicolumn{1}{c|}{0.2681}       & \multicolumn{1}{c|}{0.4516}       & \multicolumn{1}{c|}{0.6472}       & \multicolumn{1}{c|}{0.8184}       & \multicolumn{1}{c|}{0.9272}       & 0.9803       \\ \cline{2-12} \rule{0pt}{7pt}
                                        & 10                                   & \multicolumn{1}{c|}{0.0066}       & \multicolumn{1}{c|}{0.0113}       & \multicolumn{1}{c|}{0.0191}       & \multicolumn{1}{c|}{0.0381}       & \multicolumn{1}{c|}{0.0608}       & \multicolumn{1}{c|}{0.1043}       & \multicolumn{1}{c|}{0.1593}       & \multicolumn{1}{c|}{0.2398}       & \multicolumn{1}{c|}{0.3253}       & 0.4320       \\ \hline  \rule{0pt}{7pt}
  \multirow{5}{*}{C\&W}                 & 200                                  & \multicolumn{1}{c|}{0.0600}       & \multicolumn{1}{c|}{0.3450}       & \multicolumn{1}{c|}{0.7875}       & \multicolumn{1}{c|}{0.9800}       & \multicolumn{1}{c|}{1.0000}       & \multicolumn{1}{c|}{1.0000}       & \multicolumn{1}{c|}{1.0000}       & \multicolumn{1}{c|}{1.0000}       & \multicolumn{1}{c|}{1.0000}       & 1.0000       \\ \cline{2-12} \rule{0pt}{7pt}
                                        & 100                                  & \multicolumn{1}{c|}{0.0250}       & \multicolumn{1}{c|}{0.1225}       & \multicolumn{1}{c|}{0.3875}       & \multicolumn{1}{c|}{0.7313}       & \multicolumn{1}{c|}{0.9338}       & \multicolumn{1}{c|}{0.9950}       & \multicolumn{1}{c|}{1.0000}       & \multicolumn{1}{c|}{1.0000}       & \multicolumn{1}{c|}{1.0000}       & 1.0000       \\ \cline{2-12} \rule{0pt}{7pt}
                                        & 50                                   & \multicolumn{1}{c|}{0.0150}       & \multicolumn{1}{c|}{0.0531}       & \multicolumn{1}{c|}{0.1438}       & \multicolumn{1}{c|}{0.3119}       & \multicolumn{1}{c|}{0.5856}       & \multicolumn{1}{c|}{0.8038}       & \multicolumn{1}{c|}{0.9519}       & \multicolumn{1}{c|}{0.9931}       & \multicolumn{1}{c|}{0.9994}       & 1.0000       \\ \cline{2-12} \rule{0pt}{7pt}
                                        & 25                                   & \multicolumn{1}{c|}{0.0069}       & \multicolumn{1}{c|}{0.0238}       & \multicolumn{1}{c|}{0.0478}       & \multicolumn{1}{c|}{0.1019}       & \multicolumn{1}{c|}{0.1981}       & \multicolumn{1}{c|}{0.3481}       & \multicolumn{1}{c|}{0.5525}       & \multicolumn{1}{c|}{0.7534}       & \multicolumn{1}{c|}{0.9163}       & 0.9863       \\ \cline{2-12} \rule{0pt}{7pt}
                                        & 10                                   & \multicolumn{1}{c|}{0.0055}       & \multicolumn{1}{c|}{0.0058}       & \multicolumn{1}{c|}{0.0085}       & \multicolumn{1}{c|}{0.0161}       & \multicolumn{1}{c|}{0.0266}       & \multicolumn{1}{c|}{0.0479}       & \multicolumn{1}{c|}{0.0780}       & \multicolumn{1}{c|}{0.1338}       & \multicolumn{1}{c|}{0.2075}       & 0.3304       \\ \hline  \rule{0pt}{7pt}
  \multirow{5}{*}{LBFGS}                & 200                                  & \multicolumn{1}{c|}{0.0375}       & \multicolumn{1}{c|}{0.2750}       & \multicolumn{1}{c|}{0.8750}       & \multicolumn{1}{c|}{0.9950}       & \multicolumn{1}{c|}{1.0000}       & \multicolumn{1}{c|}{1.0000}       & \multicolumn{1}{c|}{1.0000}       & \multicolumn{1}{c|}{1.0000}       & \multicolumn{1}{c|}{1.0000}       & 1.0000       \\ \cline{2-12} \rule{0pt}{7pt}
                                        & 100                                  & \multicolumn{1}{c|}{0.0225}       & \multicolumn{1}{c|}{0.0763}       & \multicolumn{1}{c|}{0.3775}       & \multicolumn{1}{c|}{0.7725}       & \multicolumn{1}{c|}{0.9738}       & \multicolumn{1}{c|}{0.9975}       & \multicolumn{1}{c|}{1.0000}       & \multicolumn{1}{c|}{1.0000}       & \multicolumn{1}{c|}{1.0000}       & 1.0000       \\ \cline{2-12} \rule{0pt}{7pt}
                                        & 50                                   & \multicolumn{1}{c|}{0.0200}       & \multicolumn{1}{c|}{0.0269}       & \multicolumn{1}{c|}{0.0925}       & \multicolumn{1}{c|}{0.2888}       & \multicolumn{1}{c|}{0.5725}       & \multicolumn{1}{c|}{0.7981}       & \multicolumn{1}{c|}{0.9375}       & \multicolumn{1}{c|}{0.9850}       & \multicolumn{1}{c|}{1.0000}       & 1.0000       \\ \cline{2-12} \rule{0pt}{7pt}
                                        & 25                                   & \multicolumn{1}{c|}{0.0122}       & \multicolumn{1}{c|}{0.0188}       & \multicolumn{1}{c|}{0.0394}       & \multicolumn{1}{c|}{0.0988}       & \multicolumn{1}{c|}{0.1991}       & \multicolumn{1}{c|}{0.3528}       & \multicolumn{1}{c|}{0.5341}       & \multicolumn{1}{c|}{0.6994}       & \multicolumn{1}{c|}{0.8406}       & 0.9281       \\ \cline{2-12} \rule{0pt}{7pt}
                                        & 10                                   & \multicolumn{1}{c|}{0.0071}       & \multicolumn{1}{c|}{0.0091}       & \multicolumn{1}{c|}{0.0171}       & \multicolumn{1}{c|}{0.0293}       & \multicolumn{1}{c|}{0.0465}       & \multicolumn{1}{c|}{0.0741}       & \multicolumn{1}{c|}{0.1141}       & \multicolumn{1}{c|}{0.1689}       & \multicolumn{1}{c|}{0.2468}       & 0.3289       \\ \bottomrule
  \end{tabular}
  \label{tab: e2}  
  \end{table*}

\subsubsection{Impact of Attack Frequency}
Generally, attackers may not keep launching attacks overall timeslots. We are interested in how the frequency of the attacks affects the results of our detection method. We define the metric of attack occurrence ratio (i.e., the probability that a timeslot will be attacked) to measure the attack intensity. Fig.~\ref{fig: e7_3} shows the detection rates of the DDB-based methods with the attack occurrence ratios going from 0.1 to 1. 

It can be seen from Fig.~\ref{fig: e7_3} that the detection rates of all four methods improve as the attack occurrence ratio becomes larger. The detection rate of each method is low when the attack occurrence ratio is low. In such a scenario, the fusion center's decision will not be substantially affected because the attack happens rarely. When the attack occurrence ratio increases to 1, the detection performance gradually approaches 100\% for each method. All four methods achieve similar detection performance while our method holds slight advantages over DeepFool, C\&W, and LBFGS. 

We also measure the detection rates under different test group sizes in Table~\ref{tab: e2}. We can see from the table that a larger group size gives a better detection rate (but delays a detection decision). The detection rate in our method is 96\% when the attack occurrence ratio is 30\% with a group size of 200. But when the group size is 10, the detection rate is only 1.43\%. This is due to the fact that the larger number of test data can more accurately reflect the distribution of data in the test group. As a result, we need to increase the test group size in order to detect a low-frequency attack.

\begin{figure*}[htbp]
    \centering
    \begin{minipage}{0.32\linewidth}
      \centering
      \includegraphics[width=0.9\linewidth]{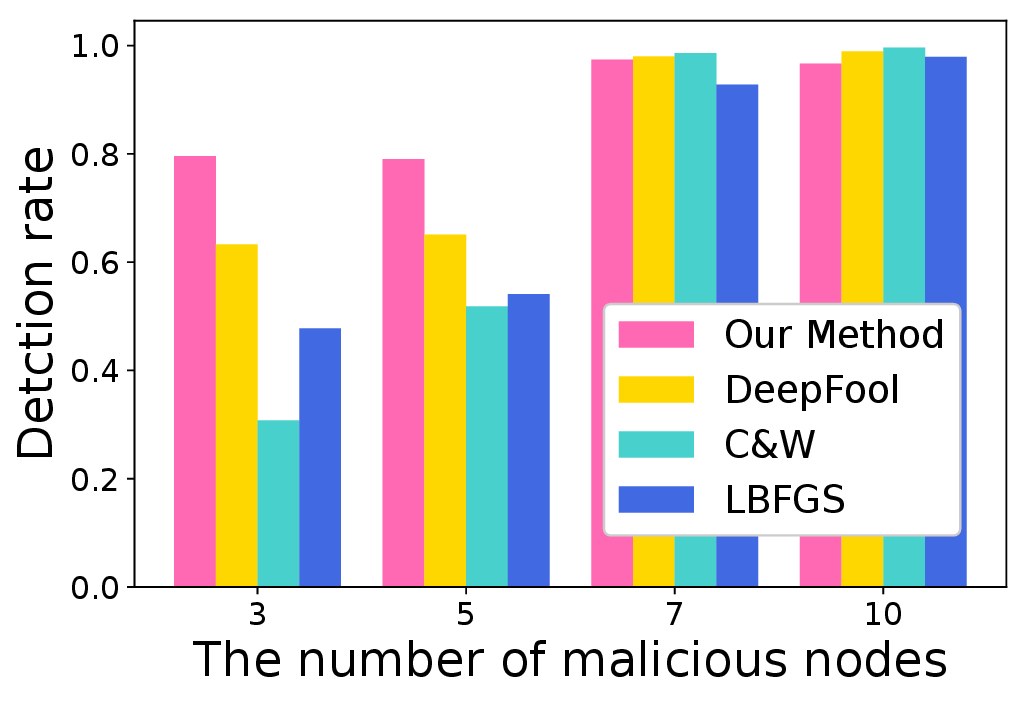}
      \vspace{-0.25cm}
      \caption{Detection rates under the different numbers of malicious nodes.}
      \label{fig: e2}
    \end{minipage}
    \centering
    \begin{minipage}{0.32\linewidth}
      \centering
      \includegraphics[width=0.9\linewidth]{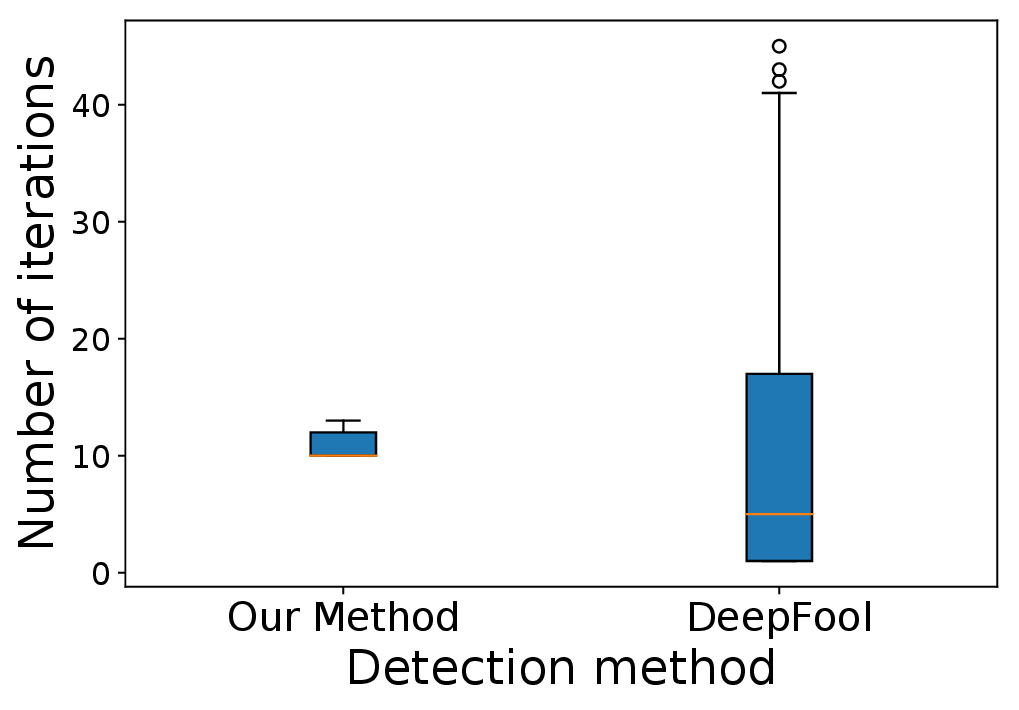}
      \vspace{-0.25cm}
      \caption{Comparison of the number of iterations between our method and DeepFool.}
      \label{fig: e3}
    \end{minipage}
    \centering
    \begin{minipage}{0.32\linewidth}
      \centering
      \includegraphics[width=0.9\linewidth]{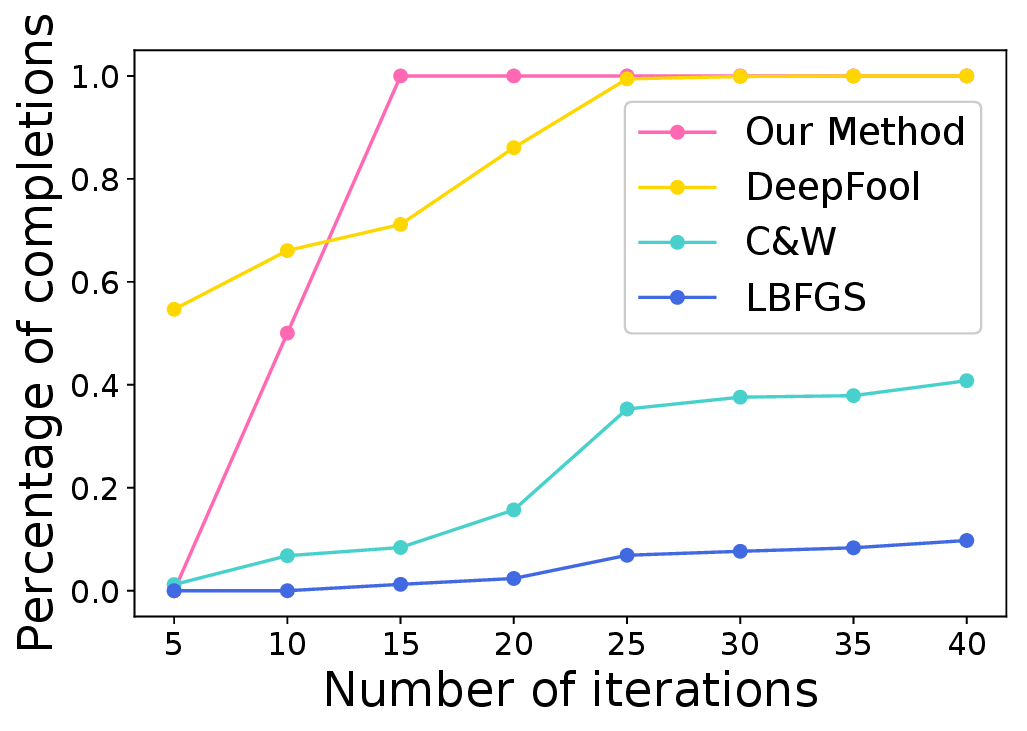}
      \vspace{-0.25cm}
      \caption{Percentages of calculations that have been completed in different given numbers of iterations.}
      \label{fig: e3_2}
    \end{minipage}
\end{figure*}

\subsubsection{Impact of Number of Malicious Nodes}
We evaluate the impact of the number of malicious nodes $m$ on the attack detection performance. In particular, we set $m = 3, 5, 7, 10$ and measure the attack detection rates in Fig.~\ref{fig: e2}.

From Fig.~\ref{fig: e2}, we observe that as $m$ increases, its influence on the DDB becomes greater, leading to a more obvious DDB distribution difference and a larger detection rate. For example, we can see when $m = 3$, the detection rate of our method is 80\%; and when $m = 10$, the detection rate becomes 98\%. In addition, we also note that even DeepFool, C\&W, and LBFGS have worse performance than our method when $m$ is small and gradually catches up with the detection rate when $m$ increases. For attackers, a large number of malicious nodes make it easier for attackers to change the value of $x$ to generate adversarial example $x'$, a significant change in $x$ also means that the DDB changes are more significant and thus easier to detect.


\subsubsection{Comparison of Time Complexity}
The evaluation results show that our DDB method generally has the sample attack detection performance, in terms of detection rate and false alarm, with DeepFool, LBFGS, and C\&W. We compare their run-time efficiency for the comparison of complexity. As all the methods are iteration-based, we use two metrics (i) the number of iterations required and (ii) the average time needed for each iteration for the complexity evaluation. We implemented DeepFool, C\&W, and LBFGS based on CleverHans V3.1.0 \cite{papernot2016technical}. 

Fig.~\ref{fig: e3} boxplots the numbers of iterations of our method compared with DeepFool. We only compare these two methods as they complete with much fewer average iterations than LBFGS and C\&W methods with hundreds or even thousands of iterations.  From the figure, we can observe that DeepFool has an average number of iterations less than that of our method (8.75 vs 10.98). However, we can see from Fig.~\ref{fig: e3} that DeepFool, due to its interactive nature, has a large variance in the number of iterations. The cumulative distribution function of the number of iterations required for each method is shown in Fig.~\ref{fig: e3_2}. It is seen from the figure that at the number of iterations of 5, DeepFool performs best with 54\% calculations completed; while our method, C\&W, and LBFGS do not have any completed calculations. When the number of iterations becomes 15, our method completes more than 99\% of the calculations, leading to a better computational complexity than the other three methods (DeepFool: 71\%, C\&W: 8\%, and LBFGS: 2\%).

\begin{table}
  \centering
  \caption{Time comparisons among different methods.}
  \begin{tabular}{ll}\toprule
  Methodology           & Time Consumption/Iteration   \\ \midrule
  Our Method &  2.41 ms   \\
  DeepFool   &  5.26 ms   \\
  C\&W       &  6.71 ms   \\
  LBFGS      &  5.85 ms   \\\bottomrule
  \label{tab: e3}
  \end{tabular}
  \vspace{-0.3cm}
  \end{table}  

Table~\ref{tab: e3} shows the average time needed for each method to complete one iteration. We can see that our method has the shortest time consumption with 2.41 microseconds per iteration. Overall, to find the DDB for one sensing data vector, our method can save, on average, 54\%, 64\%, and 59\% of the time for one iteration for DeepFool, C\&W, and LBFGS, respectively.

\begin{figure*}[htbp]    
	\begin{minipage}{0.32\linewidth}
		\centering
		\includegraphics[width=0.9\linewidth]{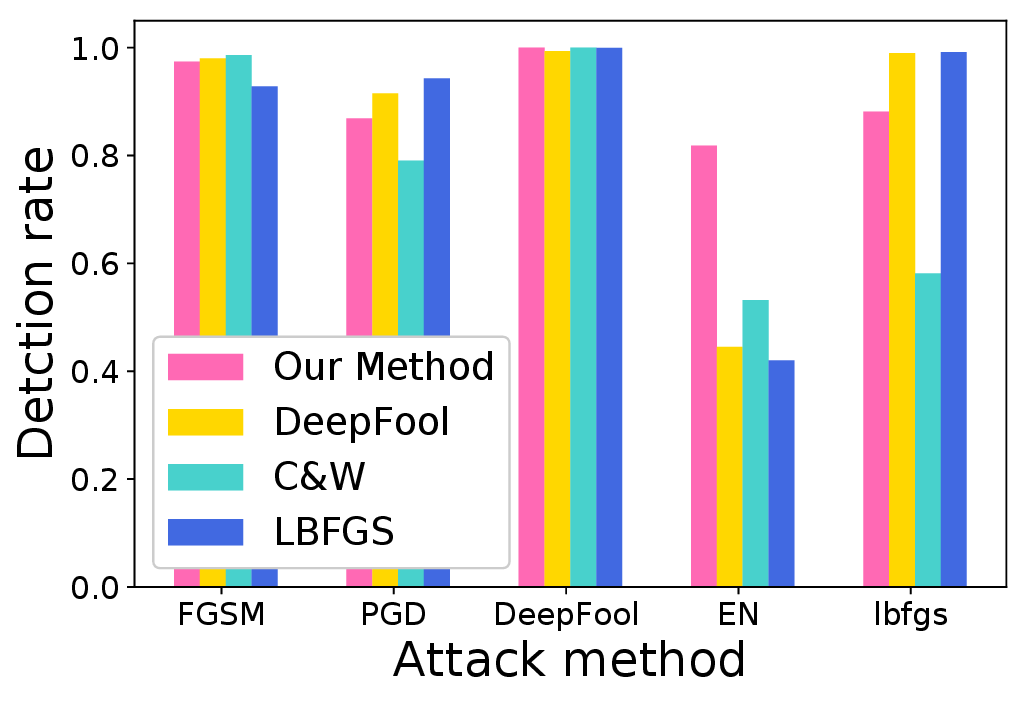}
    \vspace{-0.25cm}
		\caption{The detection rates under different attack methods.}
    \label{fig: e4}
	\end{minipage}
  \centering
  \begin{minipage}{0.32\linewidth}
		\centering
		\includegraphics[width=0.9\linewidth]{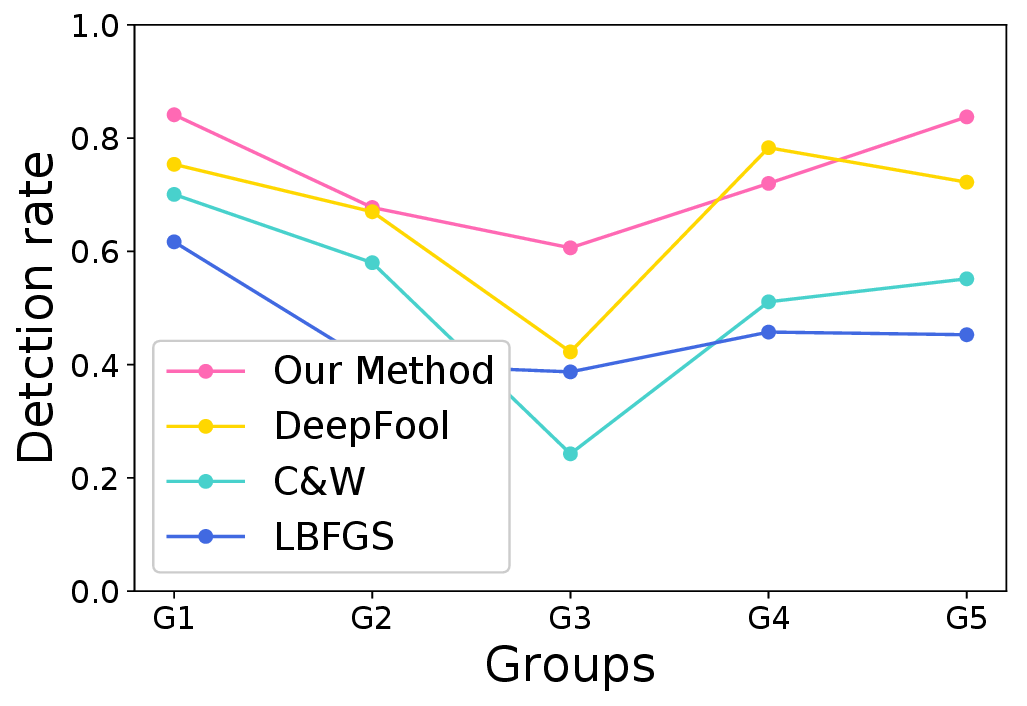}
    \vspace{-0.25cm}
		\caption{The detection rates under the different locations of malicious nodes.}
    \label{fig: e5}
	\end{minipage}
  \centering
	\begin{minipage}{0.32\linewidth}
		\centering
		\includegraphics[width=0.95\linewidth]{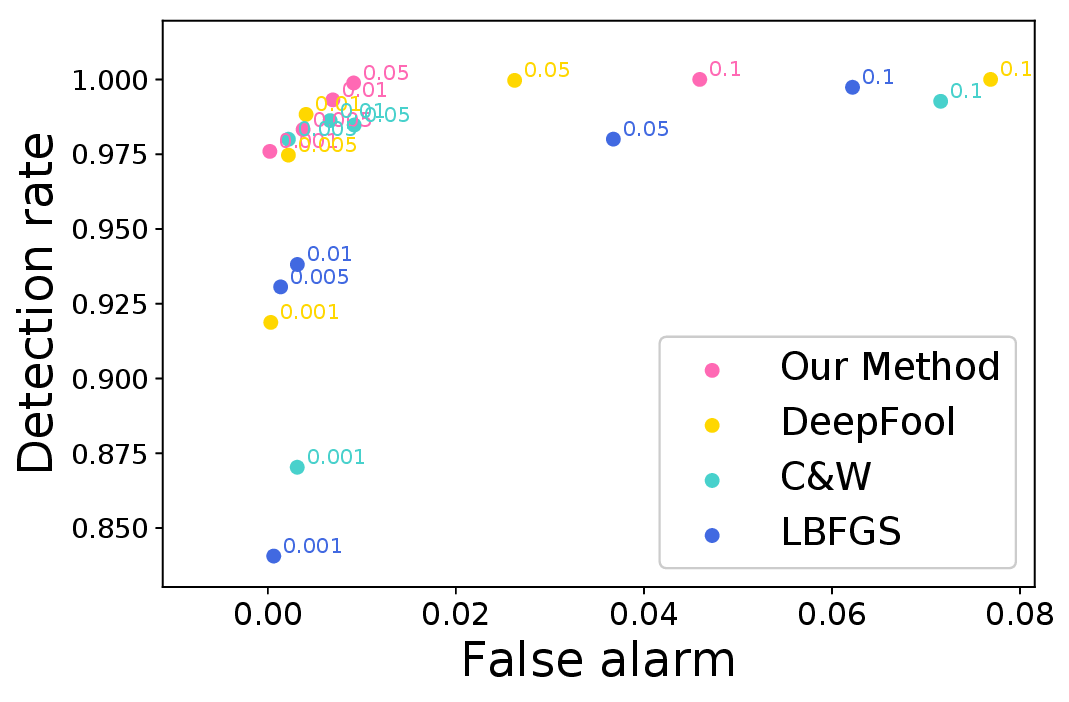}
    \vspace{-0.25cm}
		\caption{The detection rates under the different thresholds in the K-S test.}
    \label{fig: e6}
	\end{minipage}  
\end{figure*}

\subsubsection{Detection Performance under Attack Generation Methods}
Next, we compare the performance of our defense strategies in different attack approaches that the attacker may use for adversarial sensing result generation. These methods are: Fast gradient sign method (FGSM)\cite{goodfellow2014explaining}, Projected Gradient Descent (PGD)\cite{madry2017towards}, DeepFool\cite{moosavi2016deepfool}, Elastic-Net method (EN)\cite{chen2018ead} and L-BFGS\cite{szegedy2013intriguing} in Deep Neural Network, which is implemented based on CleverHans V2.1.0\cite{papernot2016technical}. The detection rate and false alarm of defending against these different attacks are shown in Fig.~\ref{fig: e4}. DDB-statistic based detection method shows good performance for most attack methods. Even if the results show that our method performs differently under different attack methods, it achieves at least 80\% detection rate and is overall better than the other three methods.

In theory, no matter which adversarial sample generation method is used, the trend of the data points will be close to the direction of the decision boundary. Our DDB-statistic based detection method can effectively detect attacks when facing different approaches adopted by the attacker, but the performance is still not ideal under some attack methods (e.g., EN). A potential reason can be that the EN method has a very low attack success rate of 8.67\% when creating adversarial example $x'$, which means that $x'$ changes less compared to $x$, which makes it difficult to detect with the DDB method. The attack success rates of using FGSM, PGD, DeepFool, and LBFGS methods are 42.37\%, 26.84\%, 57.12\%, and 31.53\%, respectively.

\subsubsection{Impact of Locations of Malicious Nodes}
Although we test different numbers of malicious nodes in our experiment, we still need to consider whether a given number of malicious nodes from different locations will affect the effectiveness of the defense. We choose 4 different nodes among the total of 20 nodes to form 5 groups of differently located malicious nodes. The detection rates are shown in Fig.~\ref{fig: e5}. It can be seen from Fig.~\ref{fig: e5} that malicious nodes in different locations will have different impacts on the attack detection performance. We find that malicious nodes have distinct attack success rates at different locations. The higher the attacker's success rate, the higher the attack detection rate. For example,  If we place malicious nodes at Location Group 1, they have the highest attack success rate (thereby causing the most damage to the network) and the DDB-based attack detection rate is also the highest in this case.  

In all methods shown in Fig.~\ref{fig: e5}, our method leads to the highest detection rate in most of the location groups, showing that it is overall more efficient than other distance methods. 

\subsubsection{Impact of Different Detection Thresholds}
We also evaluate the impact of setting the threshold $\alpha$ of $P_\text{value}$ in the K-S test on the attack detection performance. We choose 5 values of $\alpha$ in the experiment and show the results, pairing both the detection rate and false alarm, in Fig.~\ref{fig: e6}. It is noted from the figure that when $\alpha = 0.01$ (meaning that ground truth data and sensing data are different at least with the 99\% confidence
level), the detection rate and false alarm of our method are around 99.316\% and 0.691\%, respectively. And when $\alpha = 0.001$, our method and DeepFool have the lowest false alarm which is 0.0213\%. Overall, we can see that our method has similar performance to other methods while being more computationally efficient. 

\section{Related Work}\label{Sec:RW}
In this section, we summarize existing studies that are related to the work in this paper. 

\vspace{0.00cm}
{\noindent \bf Attacks and defenses in cooperative spectrum sensing:} Cooperative spectrum sensing is widely used in wireless communication systems, particularly in cognitive radio networks. Attacks against cooperative spectrum sensing can compromise the accuracy and integrity of spectrum sensing results and disrupt network operations. Some common types of attacks on cooperative spectrum sensing include: spectrum sensing data falsification (SSDF) \cite{tang2012distributed}, collusion attacks \cite{yan2012vulnerability}, and jamming attacks\cite{salameh2018spectrum}. In this paper, our defense method mainly targets SSDF, in which malicious nodes deliberately provide false or misleading sensing data to manipulate the decision-making process \cite{chen2008robust, tang2012distributed}. To counter SSDF attacking strategies, a lot of defense methods have been proposed in the literature. However, conventional studies \cite{ren2016exploiting} assumed certain prior knowledge of attacks. It has been shown in recent attack strategies \cite{luo2020attackers, adesina2022adversarial, zheng2021primary} that leveraging adversarial machine learning can beat the conventional defense and poses a new challenge to secure cooperative spectrum sensing. Our study targets the recent adversarial machine learning based attacks and adopts the concept of DDB to efficiently detect the presence of the adversarial spectrum learning attack.

\vspace{0.00cm}
{\noindent \bf Combating adversarial examples:} It is important to note that while methods have been proposed in the machine learning community to combat adversarial attacks in data classification applications, they may not necessarily work or be efficient in the spectrum sensing scenario. Generally, a machine learning model can be made more robust against small attack perturbations by training the model on a dataset that includes perturbed versions of the original data \cite{goodfellow2014explaining, wong2020fast}, or by incorporating a regularization term in the model's loss function that penalizes large perturbations in the input data \cite{finlay2019scaleable}. These methods are generally used to combat adversarial images by empirically setting a threshold of perturbations to the original data such that the perturbations do not affect the ground truth under human judgment. However, in cooperative spectrum sensing, there is no human judgment on any sensing data and there is no ground truth regarding what the sensing result should be if training data is modified with intentional perturbations. It becomes difficult to directly apply similar ideas to make the spectrum sensing data classification model robust against adversarial attacks. The DDB-statistic based detection method has shown its efficiency in detecting adversarial spectrum learning attacks. 

\vspace{0.00cm}
{\noindent \bf Computing the DDB:} The DeepFool method \cite{moosavi2016deepfool} proposed a way to compute the DDB based on a given model by iteratively perturbing input samples and checking the classifier's output until samples are misclassified. Other methods like LBFGS \cite{szegedy2013intriguing}, and C\&W \cite{carlini2017towards} aimed to generate adversarial examples using different approximations to find the shortest distance. LBFGS and C\&W methods are based on approximations to the shortest distance, and replacing the constraint with a penalty. These methods usually require multiple iterations and gradient calculations, which are time-consuming. In this work, we comprehensively compare our DDB calculation method with these methods to show the advantage of our method for the cooperative spectrum sensing scenario. 

\section{Conclusion}\label{Sec:Con}
In this paper, we presented a novel defense method for detecting malicious sensing values in cooperative spectrum sensing. Our approach leverages the distribution of the distance from a sensing value to the decision boundary of a classification algorithm, providing an effective means of detecting adversarial attacks. The experimental results demonstrate the advantages in terms of the efficiency and effectiveness of our proposed method over existing approaches for calculating the distance to the decision boundary. By effectively detecting adversarial attacks, our method can improve the overall performance of spectrum sensing systems. 

\vspace{0.1cm}
\noindent{\bf Acknowledgement:} The work at University of South Florida was supported in part by NSF Grants 2029875 and 2044516. The work at University of Miami was supported in part by NSF Grant 2029858.


\bibliographystyle{IEEEtran}
\bibliography{references}

\end{document}